\documentclass[prd,preprint,nofootinbib]{revtex4}

\usepackage{color}
\usepackage{graphicx}
\usepackage{amsfonts}
\usepackage{latexsym}
\usepackage{amsmath}
\usepackage{amssymb}
\usepackage{epsfig}

\begin{document}

\title{
Superheavy WIMP dark matter from incomplete thermalization
}
\author{Nobuchika Okada}
 \email{okadan@ua.edu}
 \affiliation{
Department of Physics and Astronomy, 
University of Alabama, Tuscaloosa, Alabama 35487, USA}

\author{Osamu Seto}
 \email{seto@particle.sci.hokudai.ac.jp}
 \affiliation{Institute for the Advancement of Higher Education, Hokkaido University, Sapporo 060-0817, Japan}
 \affiliation{Department of Physics, Hokkaido University, Sapporo 060-0810, Japan}

%

\begin{abstract}
Although it is usually thought that a class of weakly interacting massive particle (WIMP) dark matters (DMs), which have the vector coupling with the $Z$ boson, is denied by null results of the direct DM searches, such WIMP DMs are still viable if they are superheavy with the mass of $m_{DM} \gtrsim 10^{9}$ GeV. In the future, the superheavy WIMP DMs can be searched up to $m_{DM} \simeq 10^{12}$ GeV, which corresponds to the so-called neutrino floor limit. We show that the observed abundance of $\Omega _{\text{DM}}h^{2} \simeq 0.1$ for a superheavy WIMP DM can be reproduced by a suitable reheating temperature of $T_{R} \simeq m_{DM}/29$ after inflation, if the direct inflaton decay into DM is negligible or kinematically forbidden.
\end{abstract}


\preprint{EPHOU-21-006} 

\vspace*{1cm}
\maketitle


\section{Introduction}
\label{sec1}

The nature and identity of dark matter (DM) in our Universe is one of the
prime questions in particle physics and cosmology. Possible mass of DM
candidates spans the wide range from about $10^{-22}$ eV of the so-called
fuzzy dark matter~\cite{Hu:2000ke} to a macroscopic mass of primordial
black holes~\cite{Hawking:1971ei,Carr:1974nx}.

Among various DM candidates, a weakly interacting massive particle (WIMP)
with the weak scale mass and its absolute stability guaranteed by an additional
symmetry, such as a $Z_{2}$ parity, has been regarded as a particularly
interesting candidate, because of its high predictability. The relic abundance
of a WIMP DM is determined only by its interactions with particles in the
Standard Model (SM). The WIMP DM-SM interaction strength suitable for the
observed DM abundance also predicts the magnitude of the cross section
feasible to the direct DM detection experiments. Thus, the observation
of the DM relic abundance and the direct DM search are complementary methods
to probe the WIMP DM scenario.

In the standard, thermal WIMP DM scenario, the viable mass range of DM
is limited by the perturbative unitarity of the DM annihilation cross section~\cite{Griest:1989wd}.
To reproduce the observed DM abundance of
$\Omega _{\text{DM}}h^{2} \simeq 0.1$ in the freeze out mechanism, the
annihilation cross section of about $1$ pb is required. Since the cross
section decreases as the WIMP DM mass ($m_{DM}$) increases, it becomes
impossible for $m_{DM} > {\mathcal{O}}(10)$ TeV to realize the desired magnitude
of the cross section while satisfying the unitarity bound.

In principle, we can consider a WIMP DM particle with its mass much larger
than the unitarity bound. A basic question for this case is ``how can the
observed DM relic abundance be reproduced?'' The DM annihilation cross section
satisfying the perturbative unitarity bound results in the overabundance
of DM particles. There have been attempts to make the thermal superheavy
DM scenario viable. Recent attempts include a freeze-out mechanism for
a DM particle accompanied by many degenerate states~\cite{Kim:2019udq},
a huge mass gain by a late time phase transition after freeze-out~\cite{Davoudiasl:2019xeb},
DM annihilations with DM number non-conserving
$2 \leftrightarrow 2 $ interactions~\cite{Kramer:2021hal}, the DM thermalization
with not the SM thermal plasma but a hidden sector's~\cite{Hambye:2020lvy},
and the DM freeze-out during a matter-dominated era after inflation~\cite{Dunsky:2020yhv}.

In this paper, motivated by the fact that ``Weakly Interacting'' in the
word ``WIMP'' originally and literally stands for the electroweak
$SU(2)_{L} \times U(1)_{Y}$ interaction of the SM~\cite{Lee:1977ua}, we
consider a conventional weak-interacting DM particle and no new interaction
is introduced for the WIMP DM particle. Since the gauge coupling constants
of the $SU(2)_{L} \times U(1)_{Y}$ interaction are of the order of unity,
the only way to avoid the overabundance of superheavy WIMP DM is to set
an appropriate reheating temperature $T_{R}$ after inflation lower than
$m_{DM}$. In Ref.~\cite{Choi:2019pos}, this mechanism has been applied
to axino-like particle, fermionic superpartner of ALPs called ``ALPino'',
dark matter with a small ALPs decay constant as the weak scale. In the
following, we will show that a suitable value of the reheating temperature
$T_{R} \simeq m_{DM}/29$ can reproduce the observed relic abundance for
a superheavy WIMP DM.\footnote{While we investigate freeze-in scenario in
this paper, we note that impacts of a low reheating temperature on WIMP
DM freeze out have been studied in Ref.~\cite{Roszkowski:2014lga}.}

\section{A simple example: vector-like $SU(2)_L$ doublet fermion DM}
\label{sec:vlfdm}

\begin{table}[h]
	\centering
	\begin{tabular}{|c|cc|} \hline	
		 &  $SU(2)_L$ & $U(1)_Y$   \\ \hline
		$\psi_L$ &  $\mathbf{2}$ & $-\frac{1}{2}$  \\
		$\psi_R$ &  $\mathbf{2}$ & $-\frac{1}{2}$  \\ \hline
	\end{tabular}
\caption{
A simple example of the superheavy WIMP DM: a vector-like pair of $SU(2)_L$ doublet fermions.
}
\label{table1}
\end{table}
As a simple example of the superheavy WIMP DM scenario, let us consider
a vector-like pair of the $SU(2)_{L}$ doublet fermions listed in Table~\ref{table1}.
These fermions are analogous to the Higgsinos in supersymmetric models.
The mass term of the doublet fermions is given by
\begin{align}
\mathcal{L} & \supset -m \overline{\psi}_L \psi_R + \mathrm{h.c.} ,  \\
\psi_{L/R} & = \left(
\begin{array}{l}
  \chi^0_{L/R} \\
  \chi^-_{L/R} \\
\end{array}
  \right), 
\label{b5}
\end{align}
where $m$ is the Dirac mass, which may be provided by a vacuum expectation value (VEV) 
  of a scalar field in a ultraviolet completed model. 
Assigning an odd $Z_2$ parity to $\psi_{L/R}$ while even for all the SM particles, 
  the Dirac fermion from the charge neutral components ($\chi^0_{L/R}$) is stable and hence the DM particle, 
  which we simply denote by $\chi$ in the following. 
The charged fermion $\chi^\pm$ is the $SU(2)_L$ doublet partner whose mass is degenerate 
  with the DM mass. 
After the electroweak symmetry breaking, this mass degeneracy is resolved by the mass splitting
  induced by the quantum corrections through the $W$ boson loops~\cite{Cirelli:2005uq}\footnote{
  It is also possible to introduce a vector-like pair of extra charged particles
  for generating a mass splitting at the tree-level.}.
With the mass splitting, $\chi^\pm$ decays as $\chi^\pm \rightarrow \chi + W^{\pm\, *} \rightarrow \chi + \pi^{\pm}$, 
  well before the Big Bang Nucleosynthesis.

Since the DM matter fermion $\chi$ is a Dirac fermion, its electroweak interactions involve 
  the vector-like operators. 
Such a DM particle is under very severe experimental constraints.
In the following, we discuss how to avoid them to make this DM scenario viable.

\subsection{Direct DM detection bound}
\label{subsec:direct_bound}
The prime constraint on the DM particle comes from the direct DM searches,
which excluded the fourth neutrino DM and the left-handed sneutrino DM
in the minimal supersymmetric SM many years ago~\cite{Beck:1993sb}. The
DM $\chi $ with the mass $m_{\chi }$ can scatter off nucleons through the
$Z$ boson exchange. The spin-independent (SI) cross section for the elastic
scattering is given by
\begin{equation}
\sigma_{SI} = \frac{\mu^2}{\pi}\left(\frac{Z f_p+ (A-Z) f_n}{A}\right)^2 ,
\end{equation}
 with
\begin{align}
\mu &= \frac{m_\chi m_N}{m_\chi+m_N}, \\
f_p &= \frac{g_\chi}{m_Z^2}(2 g_u +g_d), \\
f_n &= \frac{g_\chi}{m_Z^2}(g_u +2 g_d), \\
g_\chi &= \frac{1}{2}g_Z, \\
g_u &= \left(\frac{1}{4}-\frac{2}{3}s_W^2 \right)g_Z, \\
g_d &= \left(-\frac{1}{4}+\frac{1}{3}s_W^2 \right)g_Z, \\
g_Z &= \frac{2m_Z}{v} = \frac{g_2}{c_W}.
\end{align}
Here, $m_{N}$ is the mass of a target atomic nucleus $N$ of the atomic
number $Z$ and the mass number $A$, $g_{2}$ is the gauge coupling of
$SU(2)_{L}$ interaction, $m_{Z}=91.2$ GeV is the mass of the $Z$ boson,
and $s_{W} \equiv \sin \theta _{W}$ and
$c_{W} \equiv \cos \theta _{W}$ with the weak mixing angle
$\theta _{W}$. For a superheavy WIMP DM ($m_{\chi } \gg m_{N}$), the reduced
mass is $\mu \simeq m_{N}$. For the xenon target of $Z=54$ and
$A\simeq 131$, we find
\begin{equation}
\sigma_{SI} = 2.3 \times 10^{-39} \mathrm{cm}^2,
\label{Eq:SI:model}
\end{equation}
as the prediction of superheavy WIMP DM cross section. 
The latest XENON1T(2018) results~\cite{Aprile:2018dbl} can be read as
\begin{equation}
\sigma_{SI} < 8.6 \times 10^{-46} \left(\frac{m_{\chi}}{1\,\mathrm{TeV}}\right) \,\mathrm{cm}^2 ,
\label{Eq:Bound:Xenon}
\end{equation}
 for $m_{\chi} \gtrsim 100$ GeV.
By comparing Eqs.~(\ref{Eq:SI:model}) and (\ref{Eq:Bound:Xenon}), we find the current lower bound on $m_\chi$ to be 
\begin{equation}
m_{\chi} > 2.7 \times 10^9 \,\mathrm{GeV} .
\label{Eq:Bound:Mass}
\end{equation}
It is prospected that the future direct DM search experiments can reach
  the so-called neutrino floor in the ultimate sensitivity as good as~\cite{Billard:2013qya} 
\begin{equation}
\sigma_{SI} \simeq 1.76 \times 10^{-48} \left(\frac{m_{\chi}}{1\,\mathrm{TeV}}\right)\, \mathrm{cm}^2 .
\label{Eq:Bound:NuFloor}
\end{equation}
This means that the future direct DM search experiments are able to probe the DM particle 
  with the mass in the range of
\begin{equation}
   2.7 \times 10^9 < m_{\chi}\left[\mathrm{GeV}\right]< 1.3 \times 10^{12}
\label{Eq:Bound:MassfromNu}
\end{equation}
 from Eqs.~(\ref{Eq:SI:model}) and (\ref{Eq:Bound:MassfromNu}).

\subsection{DM relic abundance}
\label{subsec:abundance}
Let us consider how we can realize the observed DM relic density of 
  $\Omega_{\chi}h^2 \simeq 0.1$ for the superheavy WIMP DM particle $\chi$. 
For a sufficiently low reheating temperature $T_R < m_{\chi}$, the DM $\chi$ cannot get in thermal and chemical equilibrium.
However, note that some amount of  DM $\chi$ can be produced from 
  pair annihilations of the SM particles in the thermal plasma 
  with the center of mass energy $s \geq 4 m_{\chi}^2$ in the thermal spectrum.
Since the abundance of such high energetic components is Boltzmann suppressed,
  the resultant DM abundance is expected to be suppressed by a factor $ \left( e^{-m_{\chi}/T_R} \right)^2$. 
Due to this exponential suppression, we have a chance to reproduce $\Omega_{\chi}h^2 \simeq 0.1$ 
  for $m_\chi > 2.7 \times 10^9$ GeV.

In the following, we estimate the DM abundance for $T_R < m_{\chi}$.
The Boltzmann equation for the number density of DM $n_{\chi}$ is 
\begin{equation}
\frac{d n_{\chi}}{dt}+3 H n_{\chi} = \langle\sigma_\mathrm{eff} v\rangle n_\mathrm{EQ}^2 .
\end{equation}
$H$ is the cosmic expansion rate, 
 $n_\mathrm{EQ}$ is the equilibrium number density of the SM particles, 
 and the right-hand side,  
\begin{align}
\langle\sigma_\mathrm{eff} v\rangle n_\mathrm{EQ}^2 &= \sum_{i, j}^{\chi^0, \chi^-}n^i_{\mathrm{EQ}} n^j_{\mathrm{EQ}}\langle\sigma_{ij} v_{ij}(X,Y \rightarrow i,j)\rangle \nonumber \\
 &= \frac{T}{32\pi^4}\sum_{i, j}\int^{\infty}_{(m_i+m_j)^2} ds g_i g_j p_{ij} 4E_i E_j \sigma_{ij}v_{ij} K_1\left(\frac{\sqrt{s}}{T}\right) ,
\end{align}
 involves the effective thermal averaged pair creation (annihilation) cross section $\sigma$
 times relative velocity $v$ with taking co-annihilation processes into account~\cite{Edsjo:1997bg}. 
Here, $X$ and $Y$ are initial SM particle states, $T$ is the photon temperature, 
 $g_i$ is the internal degrees of freedom of $i$ particle ($4$ for the Dirac fermion DM $\chi$), and 
\begin{equation}
p_{ij} = \frac{(s-(m_i+m_j)^2)^{1/2}(s-(m_i-m_j)^2)^{1/2}}{2\sqrt{s}},
\end{equation}
  and $E_i$ are the momentum and energy in the center-of-mass frame, respectively,
  and $K_1(z)$ is the modified Bessel function of the 1st kind.
For the mass range of our interest in Eq.~(\ref{Eq:Bound:MassfromNu}), 
  there is no mass splitting between $\chi$ and $\chi^{\pm}$ 
  because the $SU(2)_L \times U(1)_Y$ gauge symmetry is restored at the time of production.
Thus, we can take $m_i=m_{\chi}$, $p_{ij}=\sqrt{s-4m_{\chi}^2}/2$.

The resultant DM relic density is given by
\begin{align}
\Omega_{\chi}h^2 &= \frac{m_{\chi}}{\rho_\mathrm{crit}/s_0}Y_0,
\label{Eq:Omegah2_Int} \\
Y_0 &= \int^{T_R}_{T_0}\frac{1}{s T H}\langle\sigma_\mathrm{eff} v\rangle n_\mathrm{EQ}^2 dT,
\end{align}
with $ (\rho_\mathrm{crit}/s_0)^{-1} = 2.8 \times 10^8/\mathrm{GeV}$, 
where $\rho_\mathrm{crit}$ is the critical density, $T_0$ is the photon temperature at present, 
 $s_0$ is the present entropy density,
 $s$ in the denominator of integrand function is the entropy density
\begin{equation}
s = \frac{2\pi^2}{45} g_* T^3,  
\end{equation}
 and the Hubble parameter $H$ in the radiation dominated era is given by
\begin{equation}
   H =\sqrt{ \frac{\pi^2}{90} g_*} \frac{T^2}{M_P}, 
\end{equation}
  with $g_* \simeq 100$ being the total relativistic degrees of freedom
  and $M_P=2.4 \times 10^{18}$ GeV being the reduced Planck mass.
In calculating $Y_0$, we note that $\sum_{i,j} (4E_i E_j \sigma_{ij}v_{ij}) \simeq N_\mathrm{mode} \, g_2^4/(4\pi) \sim 1$
  is approximately a constant, where $N_\mathrm{mode}$ is the number of the annihilation modes. 
After pulling out the constant, the non-trivial integral turns out to be   
\begin{align}
\int^{T_R}_{T_0}\frac{dT}{T^5}\int^{\infty}_{4m_{\chi}^2} ds \frac{\sqrt{s-4m_{\chi}^2}}{2}
 K_1\left(\frac{\sqrt{s}}{T}\right)  &=
\frac{\sqrt{\pi}}{2 m_{\chi}}
G_{2,4}^{4,0} \left( \frac{m_{\chi}^2}{T_R^2} \left|
\begin{array}{c}
 1,2 \\
 0,\frac{1}{2},\frac{3}{2},\frac{5}{2} \\
\end{array}
 \right. \right) \nonumber \\
 & \simeq 
\frac{\pi}{8 m_{\chi}} \left( 4 \frac{m_{\chi}}{T_R} + 5 \right) e^{-2\frac{m_{\chi}}{T_R}}
 \qquad \mathrm{for} \quad m_{\chi} \gg T_R ,
\end{align}
  where $G$ is Meijer G-function.
We then find
\begin{align}
Y_0
 \simeq  \left(\frac{2\pi^2 g_*}{45 M_P}\sqrt{\frac{\pi^2 g_*}{90}} \right)^{-1}\frac{g_i g_j}{32\pi^4} \sum_{i,j} (4E_i E_j \sigma_{ij}v_{ij}) \frac{\pi}{8 m_{\chi}} \left( 4 \frac{m_{\chi}}{T_R} + 5 \right) e^{-2\frac{m_{\chi}}{T_R}} .
\end{align}
By substituting this into Eq.~(\ref{Eq:Omegah2_Int}), we arrived at the final expression for the DM relic density. 

\begin{figure}[t]
\begin{center}
\includegraphics[scale =1.0]{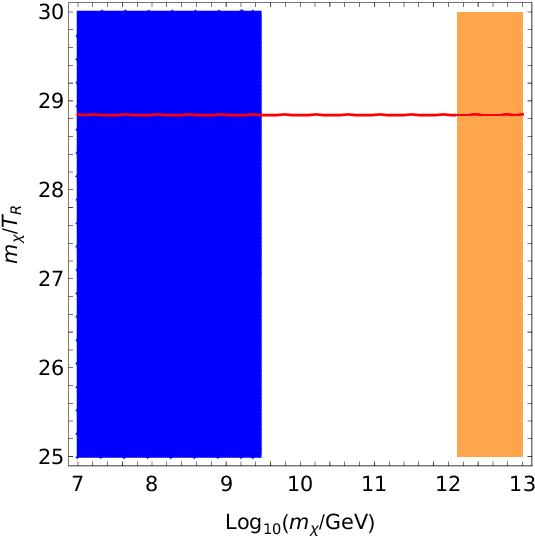} 
\end{center}
\caption{
The observed relic density is reproduced along the red line. 
The blue shaded region is excluded by the XENON1T results, while the orange shaded region 
  is beyond the search reach of the future direct DM detection experiments. 
Here, we have assumed that the superheavy WIMP DM particle is produced only 
  from the pair annihilations of the SM particles in the thermal bath.   
}
\label{fig:1}
\end{figure}

In Fig.~\ref{fig:1}, we show a relation between $m_\chi$ and $m_\chi/T_R$ 
  to reproduce the observed DM relic density of $\Omega_\chi h^2=0.1$.
We find that a suitable reheating temperature of $T_R\simeq m_\chi/29$ can reproduce $\Omega_\chi h^2=0.1$. 
The blue shaded region is excluded by the XENON1T results, while the orange shaded region 
  is beyond the search reach of the future direct DM detection experiments (beyond the neutrino floor).

\begin{figure}[t]
\begin{center}
\includegraphics[scale =1.0]{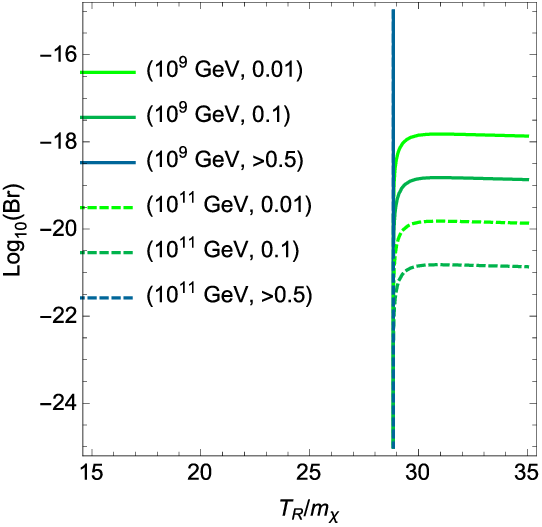} 
\end{center}
\caption{
Contours to reproduce $\Omega_{\chi} h^2 = 0.1$ 
  for various combinations of $m_{\chi}$ and $m_{\phi}$. 
Parameters in the legend denote $(m_{\chi}, m_{\chi}/m_{\phi})$.  
For $m_\chi/T_R > 29$, the DM production from the SM particles in the thermal plasma 
  is highly suppressed, and a suitable choice of the inflaton branching ratio 
  can reproduce $\Omega_{\chi} h^2 = 0.1$.  
For $m_\chi/T_R \simeq 29$,  the branching ratio should be suppressed, 
  since $\Omega_{\chi} h^2 = 0.1$ is reproduced by the thermally produced DM relic density
  as shown in Fig.~\ref{fig:1}. 
}
\label{fig:2}
\end{figure}

In the above, we have considered the creation of the superheavy WIMP DMs through $2 \rightarrow 2$ scatterings
  from the SM particles in the thermal bath.
There is another possibility that the inflaton $\phi$ with the mass $m_{\phi}$
 directly decays into a pair of DM $\chi$
  if this decay is kinematically allowed, namely $2 \,m_{\chi} < m_{\phi}$.
For the energy densities of the inflaton ($\rho_{\phi}$) and the SM particle radiation ($\rho_r$) 
 and the number density of the DM particle created by the inflation decay, 
 their evolutions are described by 
\begin{align}
&\frac{d \rho_{\phi}}{dt}+3 H \rho_{\phi} = - \Gamma \rho_{\phi} - \Gamma_{2\chi} \rho_{\phi}, \\
&\frac{d \rho_r}{dt}+4 H \rho_r = \Gamma \rho_{\phi} ,\\
&\frac{d n_{\chi}}{dt}+3 H n_{\chi} = \Gamma_{2\chi} n_{\phi},  
\label{Eq:n:chi_from_phi}
\end{align}
where $\Gamma$ is the decay width of the inflaton into the SM particles (radiation), 
  and $\Gamma_{2\chi}$ is the partial decay width for the process $\phi \rightarrow 2\chi$.
Eq.~(\ref{Eq:n:chi_from_phi}) is solved as
\begin{equation}
 a^3 n_{\chi} = \frac{\Gamma_{2\chi}}{m_{\phi}}\int^t dt' a(t')^3 \rho_{\phi}(t') ,
\end{equation}
 where we have used $\rho_{\phi} = m_{\phi} n_{\phi}$.
In the sudden decay approximation of the reheating,
 the number density of $\chi$ generated by the inflaton decay at the time of reheating is estimated as 
\begin{equation}
 n_{\chi} \simeq \frac{\mathrm{Br}}{m_{\phi}} \rho_r = \mathrm{Br}\frac{3 T_R}{4m_{\phi}} s ,
\end{equation}
where we have used the branching ratio of the process $\phi \to 2 \chi$ as
\begin{equation}
 \mathrm{Br} = \frac{\Gamma_{2\chi}}{\Gamma +\Gamma_{2\chi}} \simeq \frac{\Gamma_{2\chi}}{\Gamma} \ll 1 .
\end{equation}
By substituting $Y_0 =n_\chi/s \simeq \mathrm{Br}\frac{3 T_R}{4m_{\phi}}$ into Eq.~(\ref{Eq:Omegah2_Int}), 
  we obtain the inflaton decay contribution to the DM abundance 
  in addition to the DM abundance created from the SM thermal plasma. (See Fig.~\ref{fig:2}.)

%

Combining the DM productions from the SM particles in the thermal plasma 
   and the inflaton decay, we show the inflaton branching ratio to the DM particle pair 
   as a function of  $m_\chi/T_R$ along which $\Omega_\chi h^2=0.1$ is reproduced. 
As we have shown in Fig.~\ref{fig:1}, for $m_\chi/T_R \simeq 29$, 
   the observed DM relic abundance is fully reproduced 
   by the DM production from the SM particle annihilation in the thermal bath, 
   so that the DM production from the inflaton decay should be suppressed. 
For $m_\chi/T_R > 29$, the DM production from the thermal plasma is suppressed, 
   $\Omega_\chi h^2=0.1$ is reproduced by a suitable choice of the inflaton branching ratio.

\section{Summary}

We have considered a WIMP DM particle which has the vector-type interaction with $Z$ boson. 
Such a WIMP DM is very severely constrained by the direct DM detection experiments 
   and the current lower bound on the DM mass is found to be $m_{DM} > 2.7 \times 10^9$ GeV. 
The prime question about the superheavy WIMP DM scenario is how to reproduce 
   the observed DM relic abundance. 
As a simple example, we have considered a vector-like pair of $SU(2)_L$ doublet fermions 
   and the Dirac fermion DM is provided as their charge neutral components. 
Solving the Boltzmann equation, we have found that for a suitable value of the reheating temperature 
  $T_R < m_{DM}$, the DM particles can be produced by the SM particles with the center of mass energy 
  $s \geq 4 m_{DM}^2$ in the thermal spectrum. 
The number density fo the DM particles produced in this way is exponentially suppressed,  
  so that the thermalization of the DM particle with the SM particles has never been completed. 
We have found that for our example model, the choice of $T_R \simeq m_{DM}/29$ 
  reproduces the observed DM relic abundance of $\Omega_\chi h^2 \simeq 0.1$. 
This is our main result in this paper.  
If $T_R < m_{DM}/29$, the thermally produced DM density is too low to be consistent with the observation. 
In this case, as it is often considered, the superheavy WIMP DM particle may be produced
  by the inflaton decay and the observed DM abundance can be reproduced by 
  a suitable branching ratio of the inflaton to the DM particles.

In addition to the vector-like pair of $SU(2)_L$ doublet fermions that we have studied in this paper, 
  we can consider a variety of superheavy WIMP DM candidates. 
In the so-called minimal DM scenario \cite{Cirelli:2005uq}, we identify the charge neutral component 
  in an $SU(2)_L \times U(1)_Y$ multiplet with nonzero hypercharge as a superheavy WIMP DM. 
We may consider a special multiplet in the context of SO(10) grand unified theories 
  as a superheavy WIMP DM. 
For example, a fermion of the ${\bf 10}$ representation under SO(10) can be stable 
  thanks to the SO(10) gauge symmetry and the Lorentz symmetry 
  (see Ref.~\cite{Ferrari:2018rey} for a list of DM candidates in the SO(10) grand unified theories). 
This {\bf 10}-plet fermion includes the vector-like pair of $SU(2)_L$ doublet fermions. 
For any superheavy WIMP DM candidates, the analysis for the direct DM detection bounds 
  and the DM relic abundance are essentially the same as those presented in this paper. 
The only difference is a group factor from the $SU(2)_L$ representation of a DM multiplet. 
A superheavy WIMP DM whose mass in the range of $10^9 \lesssim m_{DM}\left[{\rm GeV}\right] \lesssim 10^{12}$, 
  which is far beyond the energy scale of high energy colliders,  
  may be discovered by the future direct DM search experiments.

%

\section*{Acknowledgments}
This work is supported in part by the U.S. DOE Grant No.~DE-SC0012447 (N.O.),
 the Japan Society for the Promotion of Science (JSPS) KAKENHI Grants
 No.~19K03860 and No.~19H05091 and No.~19K03865 (O.S.).
 




\begin{thebibliography}{99}

\bibitem{Hu:2000ke}
W.~Hu, R.~Barkana and A.~Gruzinov,
Phys. Rev. Lett. \textbf{85}, 1158-1161 (2000).

\bibitem{Hawking:1971ei}
S.~Hawking,
Mon. Not. Roy. Astron. Soc. \textbf{152}, 75 (1971).
%
\bibitem{Carr:1974nx}
B.~J.~Carr and S.~W.~Hawking,
Mon. Not. Roy. Astron. Soc. \textbf{168}, 399-415 (1974).

\bibitem{Griest:1989wd} 
  K.~Griest and M.~Kamionkowski,
  Phys.\ Rev.\ Lett.\  {\bf 64}, 615 (1990).

\bibitem{Kim:2019udq}
H.~Kim and E.~Kuflik,
Phys. Rev. Lett. \textbf{123}, no.19, 191801 (2019).

\bibitem{Davoudiasl:2019xeb}
H.~Davoudiasl and G.~Mohlabeng,
JHEP \textbf{04}, 177 (2020).

\bibitem{Kramer:2021hal}
E.~D.~Kramer, E.~Kuflik, N.~Levi, N.~J.~Outmezguine and J.~T.~Ruderman,
Phys. Rev. Lett. \textbf{126}, no.8, 081802 (2021).

\bibitem{Hambye:2020lvy}
T.~Hambye, M.~Lucca and L.~Vanderheyden,
Phys. Lett. B \textbf{807}, 135553 (2020).

\bibitem{Dunsky:2020yhv}
D.~Dunsky, L.~J.~Hall and K.~Harigaya,
JHEP \textbf{04}, 052 (2021).



\bibitem{Lee:1977ua}
B.~W.~Lee and S.~Weinberg,
Phys. Rev. Lett. \textbf{39}, 165-168 (1977).

\bibitem{Choi:2019pos}
K.~Y.~Choi, T.~Inami, K.~Kadota, I.~Park and O.~Seto,
Phys. Dark Univ. \textbf{27}, 100460 (2020).

\bibitem{Roszkowski:2014lga}
L.~Roszkowski, S.~Trojanowski and K.~Turzy\'nski,
JHEP \textbf{11}, 146 (2014).

\bibitem{Cirelli:2005uq}
M.~Cirelli, N.~Fornengo and A.~Strumia,
Nucl. Phys. B \textbf{753}, 178-194 (2006).

\bibitem{Beck:1993sb}
M.~Beck, F.~Bensch, J.~Bockholt, G.~Heusser, H.~V.~Klapdor-Kleingrothaus, B.~Maier, F.~Petry, A.~Piepke, H.~Strecker and M.~Vollinger, \textit{et al.}
Phys. Lett. B \textbf{336}, 141-146 (1994).

\bibitem{Aprile:2018dbl}
E.~Aprile \textit{et al.} [XENON],
Phys. Rev. Lett. \textbf{121}, no.11, 111302 (2018).

\bibitem{Billard:2013qya}
J.~Billard, L.~Strigari and E.~Figueroa-Feliciano,
Phys. Rev. D \textbf{89}, no.2, 023524 (2014).

\bibitem{Edsjo:1997bg}
J.~Edsjo and P.~Gondolo,
Phys. Rev. D \textbf{56}, 1879-1894 (1997).

\bibitem{Ferrari:2018rey}
S.~Ferrari, T.~Hambye, J.~Heeck and M.~H.~G.~Tytgat,
Phys. Rev. D \textbf{99}, no.5, 055032 (2019). 


\end{thebibliography}
\end{document}